\documentclass[fleqn,12pt]{iopart}
\usepackage{iopams}
\usepackage{epsfig}
\usepackage{graphicx}
\usepackage{amssymb}
\usepackage{cite}

\newif\ifcomment
\newif\ifprint
\printtrue

\newcommand{\beq}{\begin{equation}}
\newcommand{\eeq}{\end{equation}}
\newcommand{\bea}{\begin{eqnarray}}
\newcommand{\eea}{\end{eqnarray}}
\newcommand{\bce}{\begin{center}}
\newcommand{\ece}{\end{center}}

\newcommand{ \snn }{${\rm \sqrt{s_{_{NN}}}}$ =}

\def\Np{${\rm  N_{part} }$}
\def\Npavg{${\rm \langle N_{part} \rangle}$}
\def\ep{${\rm \varepsilon_{part} }$}
\def\es{${\rm \varepsilon_{std}}$}
\def\epq{${\rm \sqrt{\langle \varepsilon_{part}^{2}} \rangle}$}
\def\epr{${\rm v_{2}/\langle \varepsilon_{\rm part} \rangle }$}
\def\lsim{\mathrel{\rlap{\lower4pt\hbox{\hskip1pt$\sim$}}
    \raise1pt\hbox{$<$}}}         
\def\gsim{\mathrel{\rlap{\lower4pt\hbox{\hskip1pt$\sim$}}
    \raise1pt\hbox{$>$}}}         

\begin{document}
\vglue -1.2cm 
\title{Elliptic Flow and Initial Eccentricity in Cu+Cu and
Au+Au Collisions at RHIC} 
\author{Rachid Nouicer for the PHOBOS
Collaboration}
\address{
\vspace{2mm}
%
%
{\footnotesize
%
%
B~Alver$^4$,
B~B~Back$^1$,
M~D~Baker$^2$,
M~Ballintijn$^4$,
D~S~Barton$^2$,
R~R~Betts$^6$,
A~A~Bickley$^7$,
R~Bindel$^7$,
W~Busza$^4$,
A~Carroll$^2$,
Z~Chai$^2$,
V~Chetluru$^6$,
M~P~Decowski$^4$,
E~Garc\'{\i}a$^6$,
N~George$^2$,
T~Gburek$^3$,
K~Gulbrandsen$^4$,
C~Halliwell$^6$,
J~Hamblen$^8$,
I~Harnarine$^6$,
M~Hauer$^2$,
C~Henderson$^4$,
D~J~Hofman$^6$,
R~S~Hollis$^6$,
R~Ho\l y\'{n}ski$^3$,
B~Holzman$^2$,
A~Iordanova$^6$,
E~Johnson$^8$,
J~L~Kane$^4$,
N~Khan$^8$,
P~Kulinich$^4$,
C~M~Kuo$^5$,
W~Li$^4$,
W~T~Lin$^5$,
C~Loizides$^4$,
S~Manly$^8$,
A~C~Mignerey$^7$,
R~Nouicer$^2$,
A~Olszewski$^3$,
R~Pak$^2$,
C~Reed$^4$,
E~Richardson$^7$,
C~Roland$^4$,
G~Roland$^4$,
J~Sagerer$^6$,
H~Seals$^2$,
I~Sedykh$^2$,
C~E~Smith$^6$,
M~A~Stankiewicz$^2$,
P~Steinberg$^2$,
G~S~F~Stephans$^4$,
A~Sukhanov$^2$,
A~Szostak$^2$,
M~B~Tonjes$^7$,
A~Trzupek$^3$,
C~Vale$^4$,
G~J~van~Nieuwenhuizen$^4$,
S~S~Vaurynovich$^4$,
R~Verdier$^4$,
G~I~Veres$^4$,
P~Walters$^8$,
E~Wenger$^4$,
D~Willhelm$^7$,
F~L~H~Wolfs$^8$,
B~Wosiek$^3$,
K~Wo\'{z}niak$^3$,
S~Wyngaardt$^2$,
B~Wys\l ouch$^4$\\
}
\vspace{2mm}
{\scriptsize
$^1$~Argonne National Laboratory, Argonne, IL 60439-4843, USA\\
$^2$~Brookhaven National Laboratory, Upton, NY 11973-5000, USA\\
$^3$~Institute of Nuclear Physics PAN, Krak\'{o}w, Poland\\
$^4$~Massachusetts Institute of Technology, Cambridge, MA 02139-4307, USA\\
$^5$~National Central University, Chung-Li, Taiwan\\
$^6$~University of Illinois at Chicago, Chicago, IL 60607-7059, USA\\
$^7$~University of Maryland, College Park, MD 20742, USA\\
$^8$~University of Rochester, Rochester, NY 14627, USA\\
}}
\ead{rachid.nouicer@bnl.gov}
\begin{abstract}We present a systematic study of elliptic flow as a 
function of centrality, pseudorapidity, transverse momentum and energy 
for Cu+Cu and Au+Au collisions from the PHOBOS experiment. New data on 
elliptic flow in Cu+Cu collisions at \snn\ 22.4 GeV are shown. Elliptic flow 
scaled by participant eccentricity is found to be similar for both systems when 
collisions with the same number of participants or the same average area density 
are compared. This similarity is observed over a wide range in pseudorapidity and 
transverse momentum, indicating that participant eccentricity is the relevant quantity 
for generating the azimuthal asymmetry leading to the observed elliptic flow.
\end{abstract}
\submitto{\JPG}
\section{Introduction}
\label{sec_introduction}
\vskip -0.3cm
\hspace*{0.5cm} The characterization of elliptic flow has proven
to be one of the most fruitful probes of the dynamics of heavy ion
collisions at RHIC. It originates from the almond shape of the overlap
zone of the collision which produces, through unequal pressure
gradients, an anisotropy in the transverse momentum
distribution~\cite{Olli}. The dominant contribution to this anisotropy
is due to elliptic flow and is measured by the second coefficient,
v$_{2}$, of the Fourier expansion of the azimuthal distribution of
produced particles. The large value of v$_{2}$ observed
experimentally in semi-central Au+Au collisions at RHIC is consistent
with non-viscous hydrodynamic expansion of quark gluon plasma (QGP)
droplets~\cite{Hirano}. A strong pseudorapidity dependence of
elliptic flow reported by PHOBOS~\cite{Back_analysis,BackCuCu, Steve} provides 
useful information for constraining
models of the full three-dimensional hydrodynamic evolution of the
system.
\begin{center}
\begin{figure}[!t]
\centering
  \begin{tabular}{cc}
    \begin{minipage}{3in}
  \vspace*{0.3cm}   
         \hspace*{-1.5cm}\includegraphics[height=.28\textheight]
		 {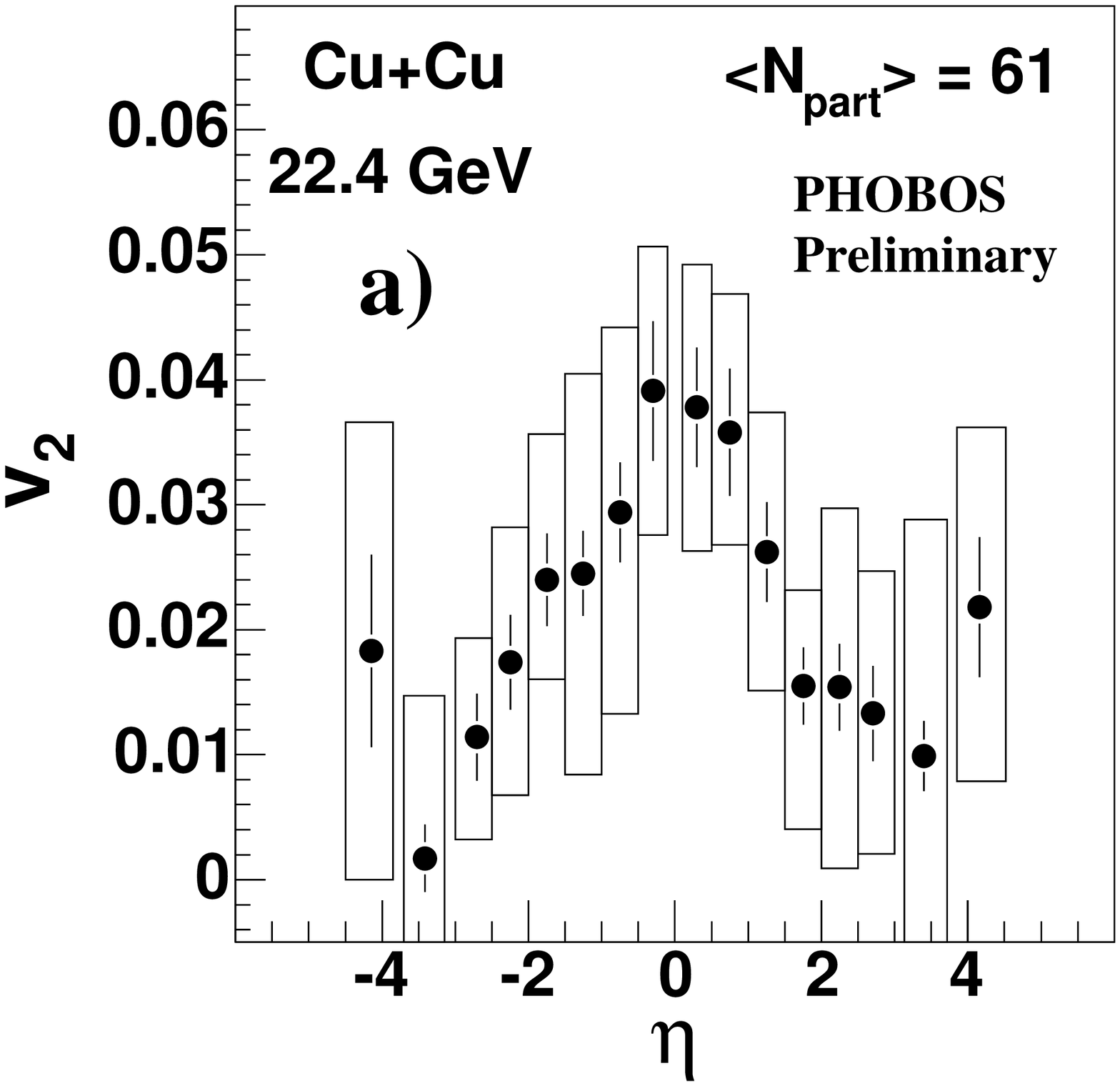}
    \end{minipage}
    &
   \begin{minipage}{3in}
       \hspace*{-2.2cm}
   \includegraphics[height=.32\textheight]
		   {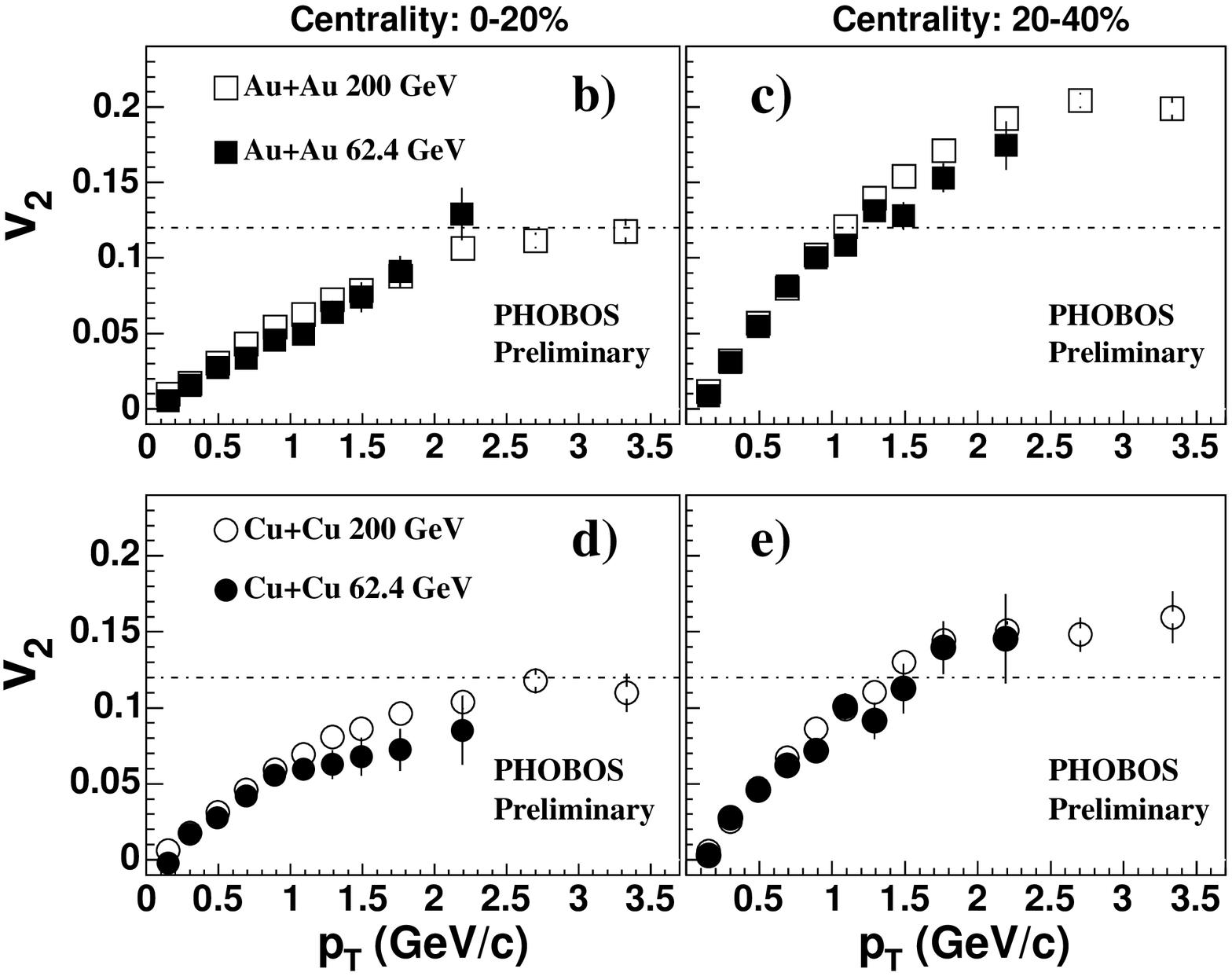}
    \end{minipage}
\end{tabular}
      \centering
\vspace*{-0.5cm}
 \caption{Panel a) elliptic flow (v$_{2}$) vs. ${\rm \eta}$ for Cu+Cu
collisions at \snn\ 22.4 GeV for centrality 0-40\% using the hit-based
method. The boxes show the systematic errors.  Panels b), c), d) and
e) v$_{2}$ vs. ${\rm p_{_{T}}}$ for Au+Au and Cu+Cu collisions at
\snn\ 62.4 and 200 GeV, b) and d) for centrality 0-20\% , c) and e)
for centrality 20-40\%. The bars in the plots represent the
statistical errors.\label{fig1}}
\end{figure}
\end{center}
\vspace*{-0.3cm}
\hspace*{0.5cm} In this paper, we present elliptic flow of charged
hadrons in Cu+Cu and Au+Au collisions at ${\rm\rm \sqrt{s_{_{NN}} }
=}$ 19.6, 22.4, 62.4 and 200 GeV as a function of pseudorapidity,
centrality and transverse momentum. The measurements of elliptic flow
in 22.4 GeV Cu+Cu collisions are shown for the first time. This work
completes our systematic study of elliptic flow measurements,
providing an extensive and precise set of experimental data for Cu+Cu
and Au+Au collisions at RHIC. Furthermore, the comparison of the data
from Cu+Cu and Au+Au collisions measured by PHOBOS experiment provides
new information on the interplay between initial state collision
geometry and elliptic flow.
\section{Results and Initial Eccentricity}
\label{sec_results}
\vskip -0.3cm
\hspace*{0.5cm} The Cu+Cu and Au+Au data presented in this work were
analyzed in the same way, using the ``hit-based'' and ``track-based''
analysis methods~\cite{Back_analysis}.  Fig.~\ref{fig1}a shows
the preliminary results of the elliptic flow signal as a function of
pseudorapidity ($\rm \eta$) in the Cu+Cu collisions at \snn\ 22.4 GeV for
0-40\% most central events. The Cu+Cu v$_{2}$ displays a strikingly
similar shape in ${\rm \eta}$ to Au+Au collisions at nearly the same
energy (19.6 GeV)~\cite{BackflowAuAu}. The strength of Cu+Cu v$_{2}$
signal is surprising in light of expectations that the smaller system
size would result in a much smaller flow signal~\cite{Chen}. The
dependence of v$_{2}$ on the transverse momentum (${\rm p_{_{T}}}$) of
charged hadrons in Au+Au and Cu+Cu collisions at \snn\ 62.4 and 200
GeV for centrality bins 0-20\% and 20-40\% are presented on
Figs.~\ref{fig1}b, 1c, 1d and 1e. We observe that for both
collision systems, the dependence of v$_{2}$ on ${\rm p_{_{T}}}$ is
similar for the two measured centrality classes. For a given system,
higher values of v$_{2}$${\rm (p_{_{T}}})$ are observed for more
peripheral collisions. 
\begin{center}
\begin{figure}[!t]
\centering
\includegraphics[height=.2\textheight]
		      {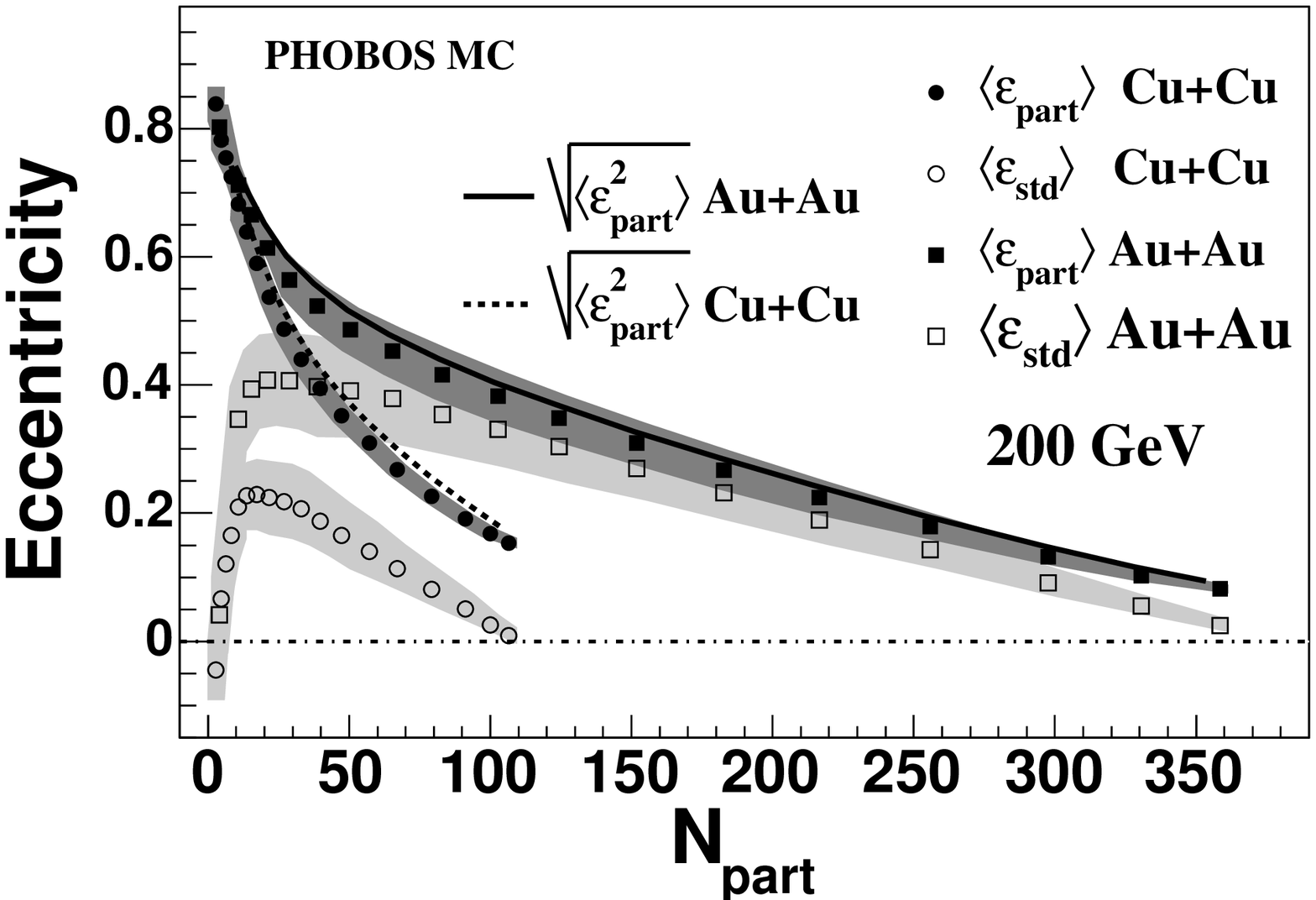}
      \centering
\vspace*{-0.5cm}
\caption{\ep\ and \es\ of the collision zone of Cu+Cu and
Au+Au at \snn 200~GeV as function of \Np\ from PHOBOS Glauber MC. The
continuous and dashed lines correspond to the RMS of \ep (\epq). The
gray bands are discussed in the text.\label{fig2}}
\end{figure}
\end{center}

\begin{center}
\begin{figure}[!t]
\centering
  \begin{tabular}{cc}
    \begin{minipage}{3in}
      \hspace*{-0.2cm}
   \includegraphics[height=.22\textheight]
		   {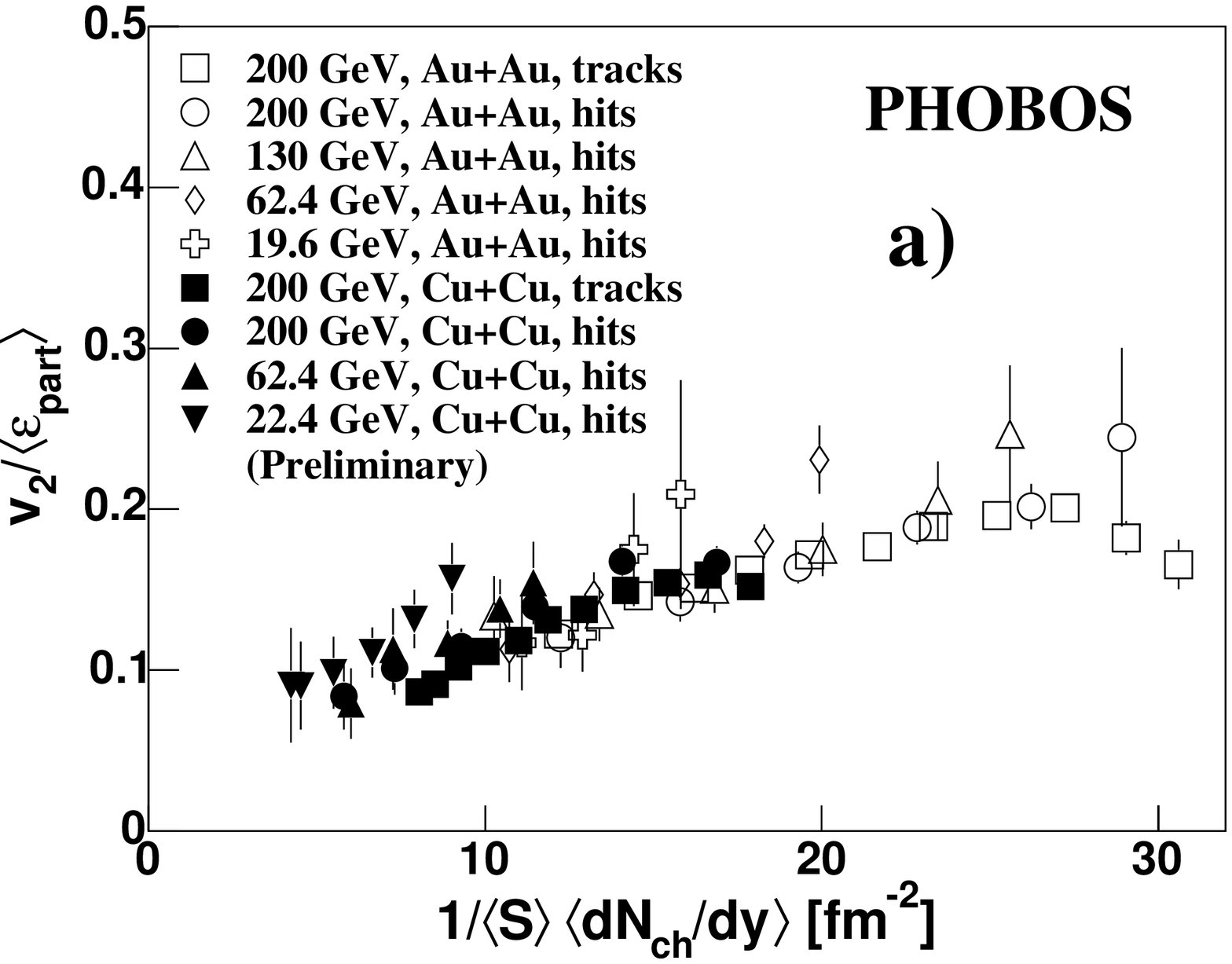}
    \end{minipage}
     &    
     \begin{minipage}{3in}
    \includegraphics[height=.22\textheight]
		    {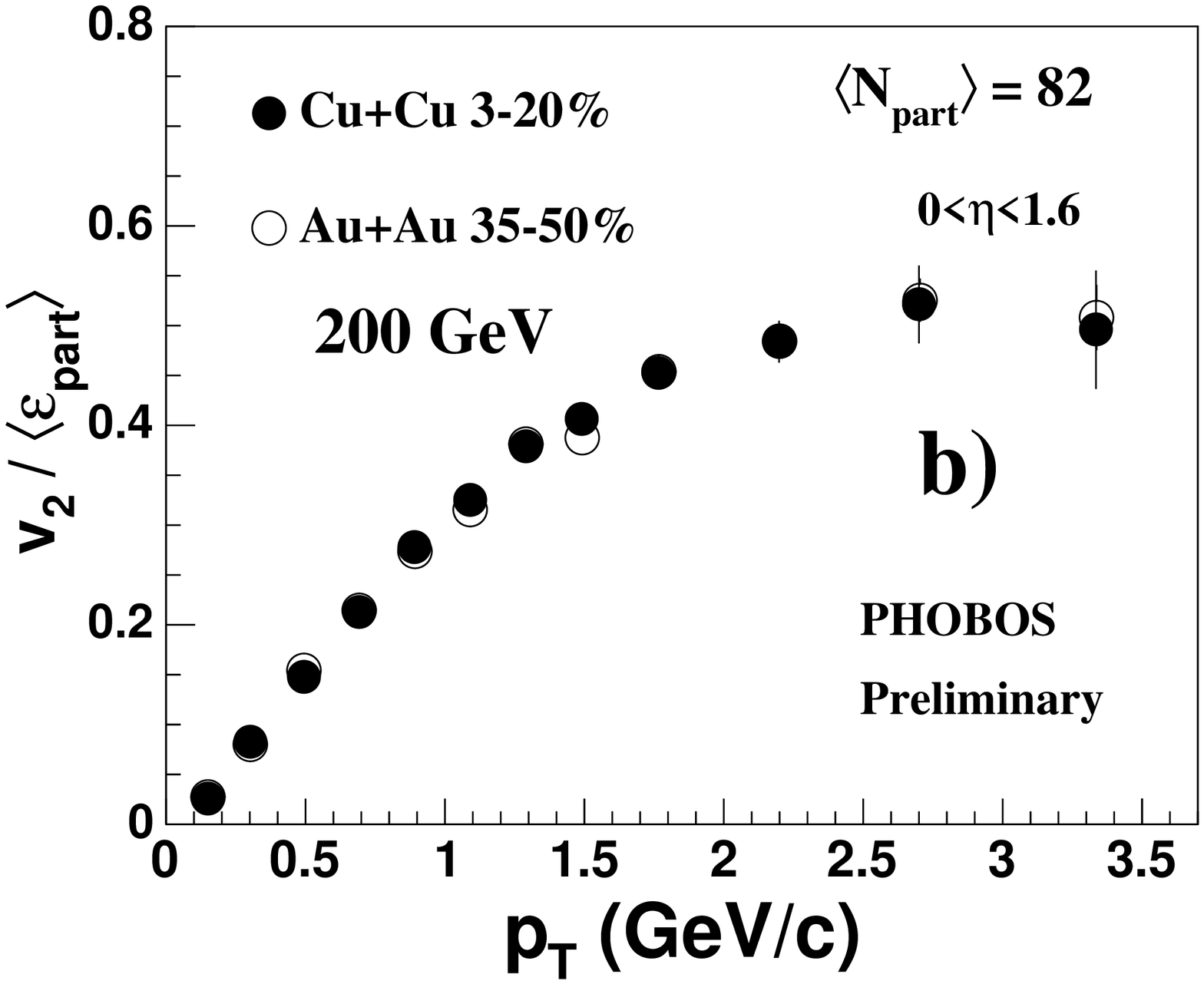}
    \end{minipage}
\end{tabular}
      \centering
\vspace*{-0.5cm}
\caption{Panel a) shows \epr\ as function of mid-rapidity (${\rm
|\eta| < 1}$) particle area density ${\rm 1/\langle S \rangle \langle
dN/dy \rangle}$ for Cu+Cu and Au+Au collisions. Panel b) \epr\ as a
function of ${\rm p_{_{T}}}$ for Cu+Cu and Au+Au collisions at 200 GeV
with the same area density (same \Npavg = 82). The bars in the plots
represent the statistical errors.
\label{fig3}}
\end{figure}
\end{center}
\begin{center}
\begin{figure}[!t]
\centering
  \begin{tabular}{ccc} 
     \begin{minipage}{3in}
    \hspace*{-0.8cm}\includegraphics[height=.16\textheight]
	    {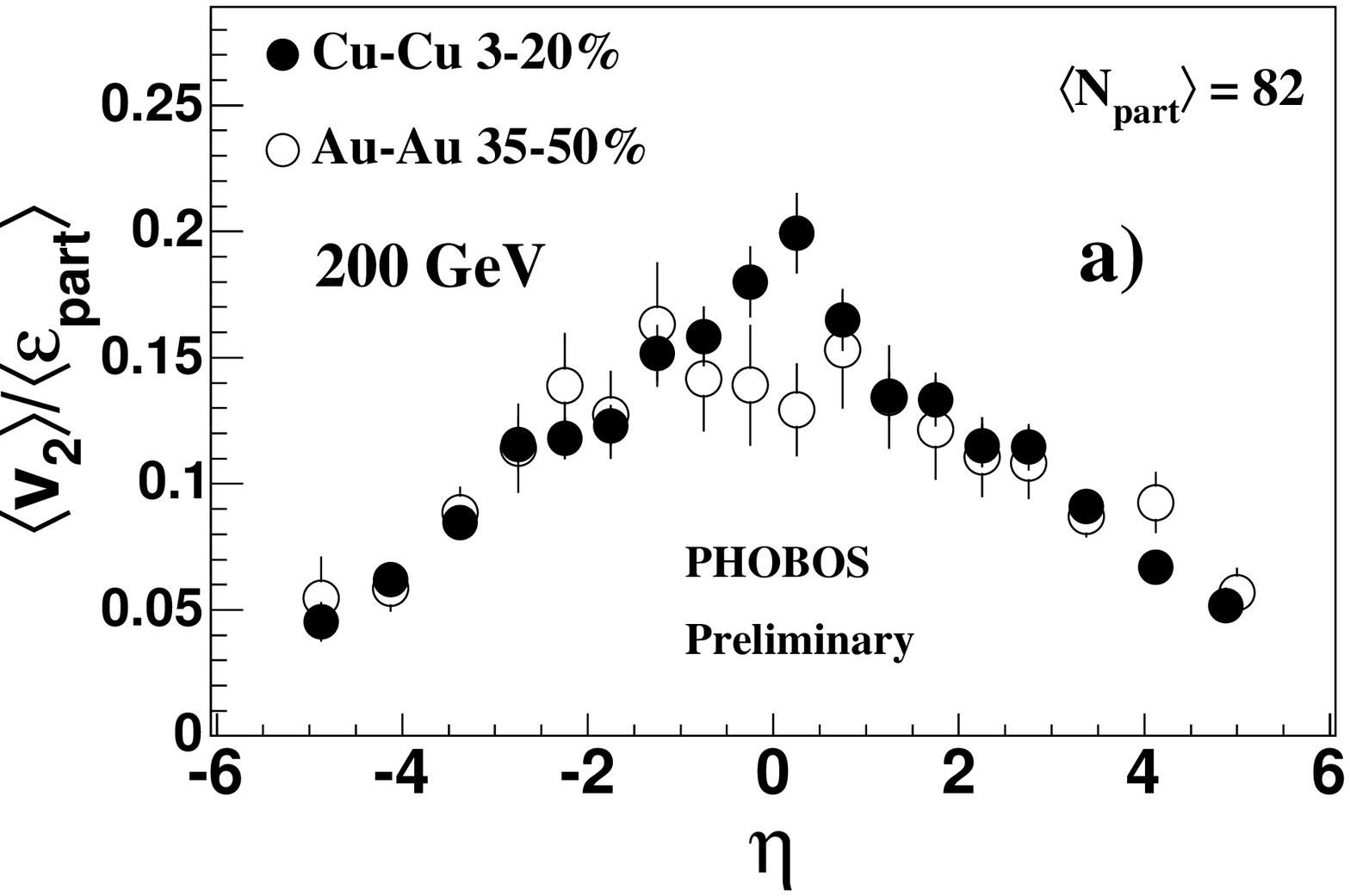}
    \end{minipage}
    & 
     \begin{minipage}{3in}
    \hspace*{-3.2cm}\includegraphics[height=.16\textheight]
	    {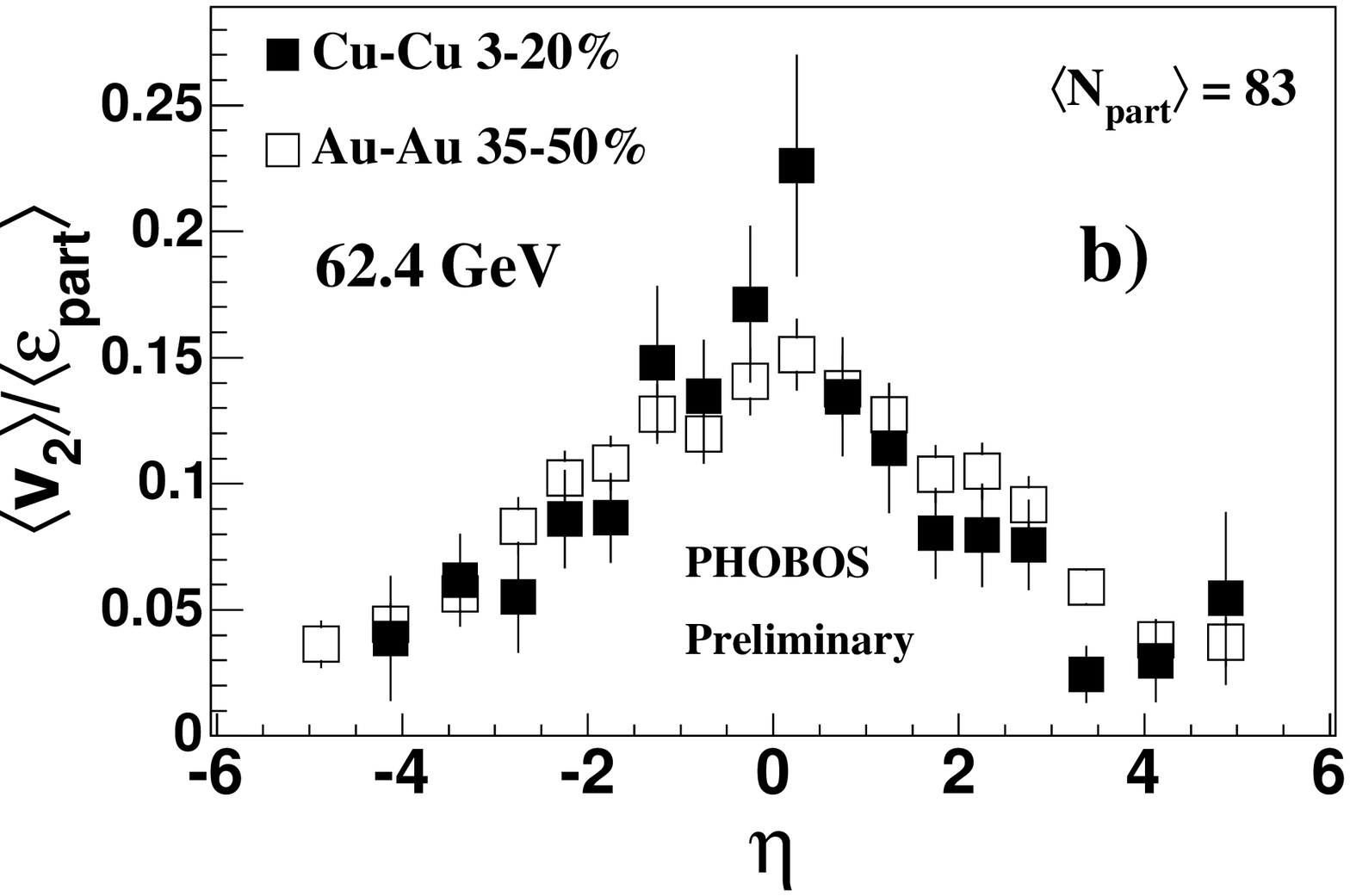}
    \end{minipage}
    &
     \begin{minipage}{3in}
    \hspace*{-5.8cm}\includegraphics[height=.16\textheight]
	    {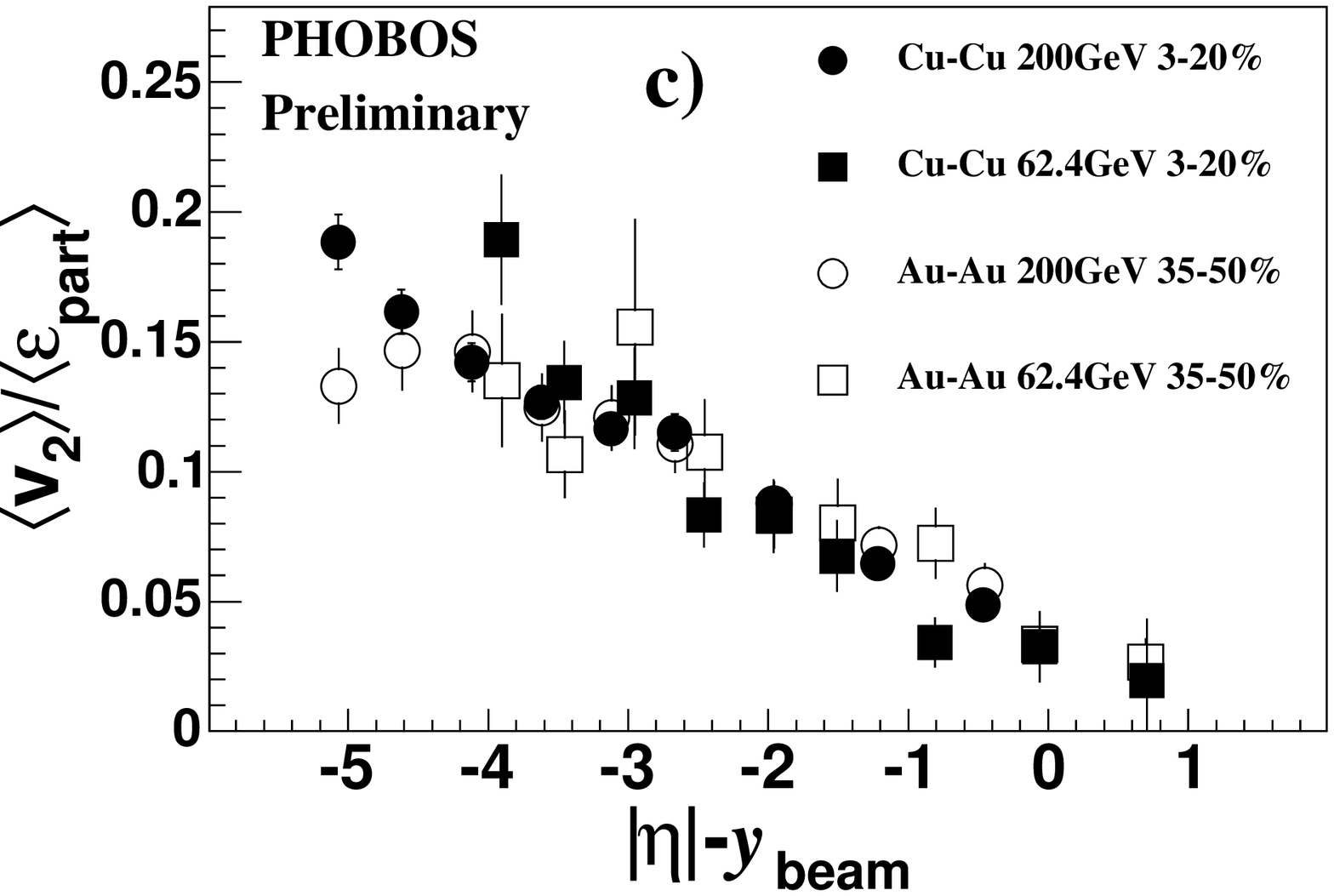}
    \end{minipage}
\end{tabular}
\vspace*{-0.5cm}
 \hspace*{-2cm}\caption{Panels a) and b) show the \epr\ as function of
${\rm \eta}$ for Au+Au (Cu+Cu) collisions with same \Npavg\ at 200 and 62.4 GeV,
respectively. Panel c) shows \epr\ as a function of ${\rm \eta' =
|\eta| - y_{beam}}$ for Cu+Cu and Au+Au collisions. The bars in the
plots represent the statistical errors.
\label{fig4}}
\end{figure}
\end{center}
\vskip -0.9cm
\hspace*{0.5cm} In order to distinguish collision dynamics from purely
geometrical effects, it has been suggested that the measured v$_{2}$ should
be scaled by the eccentricity of the nuclear overlap area~\cite{Sorge}. 
The standard definition of the eccentricity
is, $\varepsilon_{\rm std}
 =\frac{\sigma_{y}^2-\sigma_{x}^2}{\sigma_{x}^2+\sigma_{y}^2}$,
where 
$\sigma^2_{x}$~($\sigma^2_{y}$) 
are the variance of
the participant nucleon distribution projected on the $x$ ($y$) axis, taken
to be along (perpendicular to) the impact parameter direction.
\\
\hspace*{0.5cm}It has been shown that the measured v$_{2}$ in Cu+Cu
collisions at RHIC~\cite{BackCuCu,Steve} is surprisingly large even
for most central collisions, for which the average eccentricity of the
overlap region is small. The PHOBOS
collaboration has shown that for small systems
or small transverse overlap regions, event-by-event fluctuations in
the shape of the initial collision region affect the elliptic flow. Monte Carlo (MC)
Glauber studies have shown that the fluctuations in the nucleon
positions frequently create a situation where the minor axis of the
overlap ellipse of the participant nucleons is not aligned with the
impact parameter vector. To account for this effect, PHOBOS has
introduced the participant eccentricity defined
as~\cite{BackCuCu,Steve}:
${\rm \varepsilon_{\rm part} =
 \frac{\sqrt{(\sigma_{y}^2-\sigma_{x}^2)^2+4\sigma_{xy}^2}}{\sigma_{x}^2+\sigma_{y}^2},
 \label{eqeccpart}}$
where 
$\sigma_{xy}=\langle xy\rangle - \langle x\rangle\langle y\rangle$ is the covariance.
This definition accounts for the nucleon fluctuations by quantifying
the eccentricity event-by-event with respect to the overlap region of
the participants nucleons. Fig.~\ref{fig2} shows the Glauber model
calculations of \es, \ep\ and \epq\ as a function of \Np\ for Cu+Cu
and Au+Au collisions at \snn 200 GeV. The gray bands correspond to
systematic errors obtained by varying the Glauber model parameters
such as the nuclear radius, nuclear skin depth, nucleon-nucleon
inelastic cross-section and minimum nucleon separation. We observe
that \ep\ and \epq\ distributions are similar, within the small
systematic errors, for both systems. 
The importance of
the definition of eccentricity in comparing Cu+Cu and Au+Au results is
presented on Figs.~\ref{fig3} and 4, showing the eccentricity scaled
elliptic flow, v${_2/\langle \varepsilon_{part} \rangle}$, for the two
collision systems. For the comparison we selected centrality bins in Cu+Cu
and Au+Au such that \Npavg\ are matched. For such centrality bins also
the average area density, ${\rm 1/\langle S \rangle \langle dN/dy
\rangle}$, is approximately the same. We observe in Fig.~\ref{fig3}a
that the v$_{2}$ scaled by \ep\ are similar for both Cu+Cu and Au+Au
collisions at the same value of the average area density (similar
\Npavg). It should be noted that in Fig.~\ref{fig3}a which has been
introduced previousely in Ref.~\cite{Vol}, in the y-axis the
v$_{2}(\eta)$ has been converted to v$_{2}(y)$ by scaling the data by
factor 0.9 and also in the x-axis the dN/dy = 1.15 dN/d$\eta$ at
mid-rapidity region, $|\eta| < 1$. This similarity between Cu+Cu and
Au+Au collisions in Fig.~\ref{fig3}a is also observed as a function
of transverse momentum (see Fig.~\ref{fig3}b) as well as in a wide
pseudorapidity range as shown in Figs~\ref{fig4}a and 4b. Furthermore,
Fig.~\ref{fig4}c shows that the Cu+Cu and Au+Au systems at 62.4 and
200 GeV exhibit the same extended longitudinal scaling when \ep\ and
\Npavg\ are taken into consideration.
It should be noted that within
experimental errors, similar scaling properties should be observed
using \epq , as advocated in Ref.~\cite{Bhalerao2}. 
\section{Summary}
\label{sec_summary}
\vskip -0.4cm
\hspace*{0.5cm} In summary, we have performed a comprehensive
examination of the elliptic flow of charged hadrons produced in Cu+Cu
and Au+Au collisions at ${\rm \sqrt{s_{_{NN}} } =}$ 19.6, 22.4, 62.4
and 200 GeV as a function of pseudorapidity, centrality and transverse
momentum. The measurements of elliptic flow in 22.4 GeV Cu+Cu
collisions are shown for the first time. The comparison of the data
from Cu+Cu and Au+Au collisions provides new information illustrating
that the participant eccentricity is the relevant geometric quantity
for generating the azimuthal asymmetry leading to the observed
elliptic flow.

\end{document}